\documentclass[12pt,notitlepage]{article}
\usepackage{tcimemo}
\usepackage{sw20mem2}

\input tcilatex
\QQQ{Language}{
American English
}

\begin{document}

\author{Yuri V. Ralchenko\thanks{%
E-mail: \textbf{fnralch@plasma-gate.weizmann.ac.il}} \\
Department of Particle Physics\\
Weizmann Institute of Science\\
Rehovot 76100 Israel\\
\\
\\
\\
\\
\\
\\
Prepared for the Discussion Group on Data Access meeting,\\
Atomic Processes in Plasmas conference, \\
January 15, 1996, San Francisco}
\title{Atomic Data and Databases on the Internet:\\
Entering 1996\thanks{%
Electronic version can be found at %
http://plasma-gate.weizmann.ac.il/\symbol{126}fnralch/app.dvi.}}
\date{\textit{Preprint WIS-96/4/Jan-PH}}
\maketitle

\begin{abstract}
In this report the current situation with availability and management of
atomic data on the Internet is reviewed.\newpage\ 
\end{abstract}

\section{Introduction}

Only 3-4 years ago the Internet was used practically only for e-mail
correspondence. Since then the exploding development of Internet systems,
mainly of the World-Wide Web\footnote{%
The glossary of some Internet terms is given in Appendix I.}, has utterly
changed the methods of scientific information exchange. Though unlike astro-
or high-energy physicists, the atomic and plasma community have not paid
enough attention to utilizing those new opportunities in practical work, the
situation with proliferation of atomic data on the Internet rapidly changes
to better.

We review here the availability of and access to the 'wired' atomic data.
This report may be thought of as the next in the chain of analogous papers 
\cite{Martin1992,Nave1994,Smith1996}. Although because of incredibly rapid
changes in on-line atomic data distribution such a review is condemned to be
out of date almost straight after completing, nevertheless, we hope that at
least those researchers, who are not too familiar with existing on-line
atomic databases, will find new information. In what follows, we will not
stress a distinction between \textit{databases} that give search and
selection options and \textit{datasets} that are mostly ASCII or PostScript
files though this should be kept in mind. Note also that availability of
mostly\textit{\ atomic}, not molecular, data will be discussed. In addition,
no evaluation of the accuracy of concrete datasets will be provided.

In Section 2 we present the main advantages the World-Wide Web gives to
atomic data users. An effort at a comprehensive list of existing databases
is given in Section 3. Two proposals aiming more easy and convenient access
to already available and new atomic data are formulated in Section 4.
Section 5 contains the concluding remarks.

\section{World-Wide Web and Database Management}

Currently, an access to the on-line atomic data is implemented via four
protocols, viz., Telnet, e-mail, FTP, and HTTP. The HTTP-based World-Wide
Web software not only supports the other protocols but also offers many
additional options. As was correctly noted in Ref. \cite{Smith1996}, the
World-Wide Web has become a \textit{de facto} standard in modern scientific
information exchange.

The most attractive feature of the WWW is its user-friendly, extremely clear
interface. While in order to work with databases using other protocols one
has firstly to look through many-page manuals, the formation of an input on
the Web can easily be accomplished already at the very first connection to
database. Besides, some non-WWW databases run under special operating
systems and/or environment that has firstly to be learned --- this
circumstance is totally avoided in the WWW since the same client (user)
software is used for all Web databases. Another very important advantage of
the Web is that its language, HTML, was initially made to support various
text formats, images, etc., which makes the output far richer than in other
systems. Though the principles of the WWW are based on free access ideology,
nevertheless, the access can be restricted by using authentication
mechanisms. Most of client (browsers) and server (HTTP daemons) software is
free for educational, academic and governmental organizations. Finally, the
progress of the World-Wide Web and related software is so fast that we can
undoubtedly expect many new impressive features and options to appear in the
very nearest future.

\section{Atomic Databases}

In this Section we alphabetically list the on-line atomic databases and
datasets with we are familiar as of January 1996. No restricted access
databases (e.g., ADAS database) are reviewed here. The related URLs are
given in parentheses after the database names, and the collections of data
with no search options have the star superscript. The summary table is given
in Appendix II where the names of persons responsible for database operation
and/or maintainance can be found. Probably, it is hardly possible to list 
\textit{all} existing on-line atomic databases so we sincerely apologize to
the authors of the missed links.

The permanently updated hypertext list of atomic data and databases on the
Internet can be found at the URL \textit{%
http://plasma-gate.weizmann.ac.il/DBfAPP.html}.

\subsection{Atomic Data for Astrophysics$^{*}$\protect\\(\textit{%
http://www.pa.uky.edu/\symbol{126}verner/atom.html})}

This collection which is located at the University of Kentucky includes both
local data sets and many links to other hosts (mainly to the CDS catalogues)
storing atomic data. The various datasets and links are conveniently sorted
by processes or characteristics, e.g., \textsf{recombination}, \textsf{%
photoionization}, \textsf{charge transfer}, \textsf{energy levels}, \textsf{%
Stark broadening}, etc. The local data are given as ASCII and/or PostScript
tables with additional description text files.

\subsection{Atomic Data for Resonance Absorption Lines$^{*}$\protect\\(%
\textit{http://www.dao.nrc.ca/\symbol{126}dcm/atomic\_data.html})}

On the WWW server of Herzberg Institute of Astrophysics, Canada, one can
find an updated version of tables published in \cite{Morton91}. The
available data are \textsf{wavelengths}, \textsf{transition probabilities}, 
\textsf{oscillator strengths}, \textsf{damping constants}. For some of A-
and f-values the estimated accuracy is presented. The data are given as
ASCII files for different elements, the finding wavelength list being
available as well.

\subsection{BIBL\ Spectral Bibliography Database\protect\\(\textit{%
http://plasma-gate.weizmann.ac.il/bibl.html})}

The BIBL database which was created and is maintained by efforts of
Department of Atomic Spectroscopy at the Institute of Spectroscopy, Troitsk,
Russia, is not a true \textit{on-line} database. The BIBL was made for use
on IBM-PC compatibles and is added to this list so long as it can be freely
downloaded from the Weizmann Plasma-Gate site. This database contains very
detailed information on more than 4000 papers in various fields of atomic
and plasma spectroscopy and has multiple options as for searching, editing
and adding the data.

\subsection{CCP7 Data Library$^{*}$\protect\\(\textit{%
ftp://ccp7.st-and.ac.uk/ccp7/})}

One of the Collaborative Computational Projects (United Kingdom), CCP7, is
devoted to the analysis of astronomical spectra. CCP7 maintains an
unrivalled software library with atomic and molecular data at the anonymous
FTP-site of the University of St. Andrews, UK. Some of data are presented as
files which can only be read with specific software while the others can be
utilized independently. This data depository contains R. L. Kurucz CD-ROM
data, line list for LTE spectrum modeling and other data.

\subsection{Centre de Donn\'{e}es astronomiques de Strasbourg\protect\\(%
\textit{http://cdsweb.u-strasbg.fr/cats/VI.html})}

Although the main aim of \textrm{CDS} is to collect astronomical data, it
contains also about 20 catalogues (datasets) of \textsf{spectroscopic} and 
\textsf{collisional} atomic data published during last 40 years. Available
are such data as energy levels and transition probabilities from NIST
compilations, atomic spectra line list, some of Kurucz data, to name a few.
Most of catalogues can be retrieved via FTP though some should be requested
only by sending an e-mail to \textit{question@simbad.u-strasbg.fr}. The
atomic data tables from major astronomical journals are available at CDS as
well (\textit{http://cdsweb.u-strasbg.fr/cats/J.html}). Note that links to
CDS atomic data (both catalogues and tables) sorted by processes or
characteristics can be found at the \textrm{Atomic Data for Astrophysics}
site (see above).

\subsection{DASGAL Bibliography Database on Atomic Line Shapes and Shifts%
\protect\\(\textit{http://www.obspm.fr/departement/dasgal/lesage/})}

This \textsf{bibliographic} database which was created at the Department of
''Astrophysique Stellaire et Galactique'' of the Observatory of Paris-Meudon
covers period from 1977 till 1992. The available references on \textsf{line
shapes} and \textsf{shifts} are the result of joint work of Observatory of
Paris-Meudon and NIST\cite{FuhrLesage93}. The database has search options by
author, year, effect (Stark, Zeeman,...), element and title. There exist
both English and French versions of this database.

\subsection{GAPHYOR Data Center\protect\\(\textit{%
http://gaphyor.lpgp.u-psud.fr/})}

The GAPHYOR (GAz-PHYsics-ORsay) database at Centre de Donn\'{e}es, Orsay,
France covers very broad range of \textsf{bibliographic data} on properties
of isolated atoms and molecules, collisions with photons, electrons and
heavy particles (atoms and molecules), and macroscopic properties of gases.
By the middle of 1995, the total number of entries had approached half a
million. The WWW GAPHYOR database is now of version 1.1 and is rapidly
advancing. New search options are being developed in addition to already
available selection of chemical element(s) and physical process or
characteristics. The quests for expert's reports can also be sent by e-mail
to \textit{gaphyor@lpgp.u-psud.fr} but this service is not free of charge.

\subsection{Harvard-Smithsonian Center for Astrophysics Databases\protect\\(%
\textit{http://cfa-www.harvard.edu/amp/data/amdata.html})}

These are three databases which cover the atomic linelists from R. L.
Kurucz's CD-ROMs No. 18 and No. 23\cite{KuruczBell95} and R. L. Kelly's
compilation of spectral lines below 2000 \AA\ \cite{Kelly87}. The newest of
these databases (Kurucz CD-ROM No. 23, European mirror is located at \textit{%
http://leanda.pmp.uni-hannover.de:9000/projekte/kurucz/sekur.html}) has many
options in configuring both input and output while for two others the search
procedures currently include only wavelength range and/or ion range
selection. Note also that some molecular data as well as the list of other
atomic and molecular databases are presented on this server.

\subsection{IAEA Atomic and Molecular Data Information System\protect\\(%
\textit{telnet://aladdin@ripcrs01.iaea.or.at})}

This database (\textrm{AMDIS}) which contains mainly collisional data is
located at the International Atomic Energy Agency, Vienna, Austria. The
IAEA\ AMDIS (don't mix with the NIFS AMDIS!)\ consists of three databases:

\begin{itemize}
\item  \textrm{ALADDIN} -- recommended and evaluated data for \textsf{atomic
and molecular collisions} and \textsf{particle-surface interactions};

\item  \textrm{AMBDS} -- Atomic+Molecular Bibliographic Data Retrieval
System;

\item  \textrm{AMBB} -- Electronic Bulletin Board with Atomic+Molecular
related news.
\end{itemize}

The \textrm{ALADDIN}, which goes from \textbf{A} \textbf{L}abelled \textbf{A}%
tomic \textbf{D}ata \textbf{In}terface, is the system adopted by the IAEA
and the Atomic Data Centre Network for the exchange of data since 1988\cite
{Hulse90}. The non-registered users may work with userid \emph{guest} but
may not save the search results into a file. In order to become a registered
user, send an e-mail to \textit{psm@ripcrs01.iaea.or.at}. In addition to the
IAEA interactive system, \textrm{ALADDIN}\ is also available as a set of
FORTRAN-77 codes and data files which can be downloaded, e.g., from
anonymous FTP-site at \textit{ftp://ripcrs01.iaea.or.at/pub/aladdin/}.

\subsection{LLNL\ Elastic Photon-Atom Scattering Database\protect\\(\textit{%
http://www-phys.llnl.gov/V\_Div/scattering/elastic.html})}

The programs and data presented in this database (Lawrence Livermore
National Laboratory) are useful for evaluating Rayleigh scattering, the
contribution to elastic scattering made by the bound electrons of an atom,
which dominates elastic scattering for most of the x-ray and low-energy
gamma-ray regimes. These data are based on the anomalous-scattering-factor
(ASF) approximation. The existing interface provides access to ASFs for any
choice of atom and photon energy, the \textsf{differential/total cross
sections} and \textsf{angular dependence of ASFs} can be generated as well.

\subsection{Masaryk University Collection of Data$^{*}$\protect\\(\textit{%
http://www.sci.muni.cz/physics/archives.htm})}

On the anonymous FTP server of Masaryk University, Czech Republic, one can
find a small set of atomic data for some of noble gas and alkaline \textit{%
atoms}. This host contains \textsf{collisional cross sections} as well as
spectroscopic data like \textsf{energies} and \textsf{transition
probabilities}.

\subsection{NIFS\ Database\protect\\(\textit{telnet://msp.nifs.ac.jp})}

This is a large databank of \textsf{collisional processes characteristics}
which is located at the National Institute for Fusion Science (NIFS),
Nagoya, Japan. Actually, it consists of a few databases under common
envelope of the Fujitsu Advanced Information Retrieval System (FAIRS), among
them:

\begin{itemize}
\item  \textrm{AMDIS} -- Atomic and Molecular Data Interactive System;

\item  \textrm{BACKS} -- Backscattering of Ions from Solids in Normal
Incidence;

\item  \textrm{CHART} -- Charge Transfer between Atoms and Ions;

\item  \textrm{SPUTY} -- Sputtering Yields of Monoatomic Solids.
\end{itemize}

Four additional databases located at NIFS are the plasma-fusion-atomic
excerpts from the well-known commercial INSPEC$^{\registered }$ database and
the ORNL\ bibliographic database (see below). The original NIFS databases
contain very detailed and extensive information on such collisional
processes as, e.g., electron impact excitation and ionization, charge
transfer, etc. As of the end of 1995, the \textrm{AMDIS} contains 4303 sets
of electron-ion excitation cross sections and 1302 sets of electron-ion
ionization cross sections. The selected data sets can be downloaded with FTP
or plotted on a screen. There is no anonymous access to the NIFS database so
in order to get a userid with a password one has to send a request to
Research Information Center, NIFS, Nagoya 464-01, Japan.

\subsection{NIST Atomic Spectroscopy Database\protect\\(\textit{%
http://aeldata.phy.nist.gov/nist\_atomic\_spectra.html)}}

This database (currently version 1.0) is a part of the NASA Astrophysics
Data System which is based purely on the World Wide Web. \textrm{ASD}
contains critically evaluated data on \textsf{energy levels}, \textsf{%
wavelengths} and \textsf{transition probabilities} and reflects the most
up-to-date state for these spectroscopic data. Currently, included are data
of various degree of completeness for 36 elements. The search options are
very basic, though full-size help is available on-line. There also exists a
telnet version of \textrm{ASD} (\textit{telnet://asd@atm.phy.nist.gov})
which contains data only for four elements.

\subsection{NIST Atomic Transition Probability Bibliographic Database\protect
\\(\textit{http://physics.nist.gov/PhysRefData/fvalbib/reffrm0.html})}

The NIST \textsf{A-value} \textsf{bibliographic} database is one of the
first WWW atomic databases. This databank (current version is already 3.0)
contains about 3000 references up to 1994 which are part of the collection
of the Data Center on Atomic Transition Probabilities at NIST. The multiple
selection criteria include element, isoelectronic sequence, author, journal,
year of publication, method of calculation/measurement, etc.

\subsection{Opacity Project TOPbase\protect\\(\textit{%
http://cdsarc.u-strasbg.fr/OP.html})}

The TOPbase\cite{Mendoza1992} is a data management system (DMS) which was
designed specially for presentation of the results from the widely-known
Opacity Project\cite{Seaton87}. This database which is located at the Centre
de Donn\'{e}es astronomiques at Strasbourg, France, contains \textsf{energy
levels}, \textsf{wavelengths}, \textsf{oscillator strengths} and \textsf{%
photoionization cross sections} for ions of many astrophysically abundant
elements. Like other DMSs, the TOPbase offers multiple options as for
selection and presentation of available data including possibilities for
graphical output. The TOPbase site can be directly reached at the host 
\textit{cdsarc.u-strasbg.fr }with \emph{topbase} as userid and \emph{Seaton+}
as password.

\subsection{ORNL Controlled Fusion Atomic Data Center Databases\protect\\(%
\textit{http://www-cfadc.phy.ornl.gov/)}}

In addition to collection of ALADDIN codes/data and a list of atomic and
plasma databases, on the CFADC home page one can find a link to categorized 
\textsf{bibliographic} database of about 30,000 references dating from 1978
to present (work on adding 30,000 references till 1978 is in progress). All
data are divided into 9 categories and tens of subcategories which embrace
practically all atomic processes and characteristics related to nuclear
fusion. User can select input parameters from the list of (sub)categories or
make a search by author name. The other refined database at the CFADC
presents experimental ionization cross sections measured in the Physics
Division at ORNL. The CFADC host serves also as a temporary location of 
\textit{Atomic Data and Nuclear Data Tables} journal and \textit{Theoretical
Atomic, Molecular, and Optical Community} WWW home pages.

\subsection{SAM Project Data$^{*}$\protect\\(\textit{%
http://aniara.gsfc.nasa.gov/sam/sam.html})}

The aim of the ''Systematic, Accurate, Multiconfiguration calculations''
project which has recently been initiated by international collaboration of
atomic theory groups is to produce, collect and distribute accurate atomic
data with special attention paid to uncertainty evaluation. The SAM results
-- \textsf{oscillator strengths}, \textsf{energy levels}, \textsf{wavelengths%
},\textsf{\ hyperfine structure parameters} -- are stored mainly as
PostScript (PS) or DVI sources of SAM collaborators papers or, in few cases,
as PS, DVI or ASCII files of data tables. It should be added that one can
find there a lot of data on intercombination and forbidden lines which are
not well presented yet in other databases.

\subsection{Uppsala University Databases\protect\\(\textit{http://xray.uu.se/%
})}

The X-Ray WWW server at the Department of Physics at Uppsala University,
Sweden, stores various information related to X-ray physics. One can find
there the COREX database of core edge (inner shell) excitation spectra of
gas phase atoms and molecules with bibliographies, the database of Henke
scattering factors\cite{Henke}, a list of electron binding energies in eV
for the elements in their natural forms, and a list of X-ray emission lines.
The first two databases feature search by keywords with graphical output
while the lists are the ASCII files. Note that Henke scattering factors can
also be found at the URLs \textit{ftp://grace.lbl.gov/pub/sf/} and \textit{%
ftp://xray1.physics.sunysb.edu/pub/henke}.

\subsection{Vienna Atomic Line Database\protect\\(\textit{%
mailto:vald@galileo.ast.univie.ac.at})}

The only atomic database which is based solely on the e-mail interface is
the Vienna Atomic Line Database (\textrm{VALD}) at the Institut f\"{u}r
Astronomie, Vienna, Austria. The \textrm{VALD} includes data on \textsf{%
atomic line parameters} and provide tools for selecting subsets of lines for
typical astrophysical applications. The data sets are extracted from
different sources and are then critically evaluated. To become a \textrm{VALD%
}\ client, send an e-mail with your full name and \textbf{all} e-mail
addresses you may use to the \textrm{VALD} administrator at \textit{%
valdadm@galileo.ast.univie.ac.at}. With registration confirmation, user
receives the \textrm{VALD} manual.

\section{Proposals}

Concerning the nearest improvement of an access to the on-line atomic data,
we would like to pay attention of the Discussion Group to the following two
proposals:

\begin{enumerate}
\item  \begin{quote}
\emph{Organization of two hosts --- in the U.S.A. and in Europe --- that
would serve as mirrors of existing databases.}
\end{quote}

\begin{quotation}
Firstly, such mirrors should greatly facilitate an access to already
available data. Practically all database host computers are not dedicated to
atomic database management alone, and scheduled and unforeseen shut-downs
due to activity not related to database management are not very uncommon.
For instance, the NIFS databases are down almost every Saturday and Sunday.
Secondly, the mirror service team might supplement the copies of non-Web
databases with WWW interface. This task seems to be neither enormous nor
unreal because the modern management tools like, e.g., PERL computing
language, proved to be very successful in such applications. Thirdly, the
mirrors may also serve as a repository of a free atomic software and even
may run simple atomic codes under WWW envelope. In addition, the data on the
mirrors might be filtered to avoid unnecessary duplication.

The operation of large WWW databases shows that middle-class workstations
with sufficient disk space would be quite suitable for storing huge amounts
of data and processing multiple simultaneous user requests.
\end{quotation}

\item  \begin{quote}
\emph{Use of the e-Prints archive (URL http://xxx.lanl.gov).}
\end{quote}

\begin{quotation}
When visiting the WWW home pages of various atomic laboratories, one can
easily notice that almost all hosts contain the copies of local group
publications in different formats. Now the natural question arises why not
to put them onto entirely automated system with different access modes, well
developed search and selection options, immediate notification of new
submissions, and possibility of free retrieval of papers? Fortunately, such
a system does exist already for about 5 years and has \textit{really} become
the primary means in distribution of ongoing scientific information in
high-energy physics, astrophysics, quantum physics, etc. Probably, the main
widespread prejudice against e-prints is the lack of refereeing which is
believed to lead to dissemination of low-quality results. No matter how
solid this argument would be -- have you never seen questionable papers in
refereed journals? -- it must be admitted that for some fields of physics
such an open distribution of research works well and has real advantages for
researchers in developed and especially undeveloped countries. It is a pity
that the Plasma Physics (started from February 1995) and Atomic, Molecular,
and Optical Physics (started from September 1995) e-prints archives still
stand out for their very low load comparing to most other archives.
\end{quotation}
\end{enumerate}

\section{Conclusion}

Although the 'wired' atomic data are becoming more easy accessible, the
existing opportunities provided by the World-Wide Web are not fully utilized
yet. We believe that the activity of this Discussion Group in coordination
of joint efforts in further development of electronic atomic databases as
well as the regular, biennial `International Conference on Atomic and
Molecular Data for Science and Technology' recently proposed by A. Dalgarno
and R. K. Janev can be extremely effective.

\section{Acknowledgments}

It is a big pleasure to thank Peter Smith for providing me with text of Ref. 
\cite{Smith1996} prior to publication, Jeffrey Fuhr and Peter Mohr for
sending some information on DASGAL and NIST databases, and Ratko Janev for
information concerning proposed new series of conferences. The critical
reading of the manuscript by Evgeny Stambulchik is highly appreciated.
Special thanks are due to Jeffrey Nash who made important notes on the
manuscript and gave this text a chance to reach the meeting. Finally, I am
very grateful to Jeffrey Nash and David Schultz for the opportunity to
present this report to the participants.

\newpage 

\textbf{Appendix I}\\

\textsc{Glossary of Internet terms}\\

\textbf{FTP}:

\begin{quote}
\textit{File Transfer Protocol} -- a client-server protocol which allows a
user on one computer to transfer files to and from another computer over the
Internet. Also the client program the user executes to transfer files.
\end{quote}

\textbf{HTML}:

\begin{quote}
\textit{HyperText Markup Language} -- hypertext document format used by the
World-Wide Web.
\end{quote}

\textbf{HTTP}:

\begin{quote}
\textit{HyperText Transfer Protocol} -- the client-server protocol used on
the World-Wide Web for the exchange of HTML documents.
\end{quote}

\textbf{Telnet}:

\begin{quote}
The Internet standard protocol for remote login.
\end{quote}

\textbf{URL}:

\begin{quote}
\textit{Uniform Resource Locator} -- a draft standard for specifying an
object on the Internet, such as a file or newsgroup. URLs are used
extensively on the World-Wide Web. They are used in HTML documents to
specify the target of a hyperlink. Examples:
\end{quote}

\begin{quotation}
\textit{http://plasma-gate.weizmann.ac.il/DBfAPP.html}

\textit{news:sci.physics.plasma}

\textit{telnet://aladdin@ripcrs01.iaea.or.at} --- here \textit{aladdin} is
the userid.

\textit{ftp://topbase:Seaton+@cdsarc.u-strasbg.fr/users} --- here \textit{%
topbase }is the userid and \textit{Seaton+} is the password.
\end{quotation}

\textbf{WWW}:

\begin{quote}
\textit{World-Wide Web} (also W3) -- An Internet client-server hypertext
distributed information retrieval system. On the WWW everything (documents,
menus, indices) is represented to the user as a hypertext object in HTML
format. Hypertext links refer to other documents by their URLs. These can
refer to local or remote resources accessible via FTP, Gopher, Telnet or
news, as well as those available via the HTTP protocol used to transfer
hypertext documents.
\end{quote}

\newpage\ 

\textbf{Appendix II}\\

\textsc{Atomic Databases}\\

\begin{tabular}{|c|c|c|c|c|}
\hline
\textsc{Name} & \textsc{Country} & \textsc{Contact} & \textsc{Access} & 
\textsc{Data} \\ \hline
\textrm{TOPbase} & France & \textbf{C.Mendoza} & Telnet & \textsf{%
EL,OS,TP,PH,WL} \\ \hline
\textrm{NIFS} & Japan & \textbf{H.Tawara} & Telnet & \textsf{EX,IN,CT,HP} \\ 
\hline
\textrm{NIST ASD} & U.S.A. & \textbf{D.Kelleher} & WWW,Telnet & \textsf{%
EL,OS,TP,WL} \\ \hline
\textrm{CfA Harvard} & U.S.A. & \textbf{P.L.Smith} & WWW & \textsf{%
EL,OS,TP,WL} \\ \hline
\textrm{CDS} & France & \textbf{---} & FTP & \textsf{VSC} \\ \hline
\textrm{CFADC} & U.S.A. & \textbf{D.R.Schultz} & WWW & \textsf{IN,Bibl: VSC}
\\ \hline
\textrm{GAPHYOR} & France & \textbf{J.L.Delcroix} & WWW,E-mail & \textsf{%
Bibl: VSC} \\ \hline
\textrm{IAEA AMDIS} & Austria & \textbf{R.K.Janev} & Telnet & \textsf{%
EX,IN,CT,HP,Bibl} \\ \hline
\textrm{VALD} & Austria & \textbf{F.Kupka} & E-mail & \textsf{EL,OS,TP} \\ 
\hline
\textrm{CCP7} & U.K. & \textbf{C.S.Jeffery} & FTP & \textsf{EL,WL} \\ \hline
\textrm{SAM} & U.S.A. & \textbf{T.Brage} & WWW & \textsf{EL,OS,TP,HS} \\ 
\hline
\textrm{DASGAL} & France & \textbf{A.Lesage} & WWW & \textsf{Bibl: LS,SH} \\ 
\hline
\textrm{ADA} & U.S.A. & \textbf{D.Verner} & WWW & \textsf{VSC} \\ \hline
\textrm{NIST ATPBD} & U.S.A. & \textbf{P.J.Mohr} & WWW & \textsf{Bibl: OS,TP}
\\ \hline
\textrm{ADRAL} & Canada & \textbf{D.C.Morton} & WWW & \textsf{EL,OS,TP} \\ 
\hline
\textrm{BIBL} & Russia & \textbf{A.N.Ryabtsev} & FTP & \textsf{Bibl: VSC} \\ 
\hline
\textrm{LANL EPAS} & U.S.A. & \textbf{L.Kyssel} & WWW & \textsf{SF} \\ \hline
\textrm{Uppsala U.} & Sweden & --- & WWW & \textsf{SF,WL,OS} \\ \hline
\end{tabular}
\\\\

\textbf{Data keys:}

\textsf{Bibl: XXX} - bibliography for XXX; \textsf{CT} - charge transfer
cross sections; \textsf{EL} - energy levels; \textsf{EX} - electron impact
excitation cross sections; \textsf{HP} - heavy-particles interaction cross
sections; \textsf{HS} - hyperfine structure parameters; \textsf{IN} -
electron impact ionization cross sections; \textsf{LS} - line shapes; 
\textsf{OS} - oscillator strengths; \textsf{PH} - photoionization cross
sections; \textsf{SH} - line shifts; \textsf{SF} - scattering factors; 
\textsf{TP} - transition probabilities; \textsf{VSC} - various spectroscopic
and collisional data; \textsf{WL} - wavelengths.

\end{document}